# FRANCESCO INGOLI'S ESSAY TO GALILEO: TYCHO BRAHE AND SCIENCE IN THE INQUISITION'S CONDEMNATION OF THE COPERNICAN THEORY

Christopher M. Graney

Jefferson Community & Technical College

1000 Community College Drive

Louisville, KY 40272

christopher.graney@kctcs.edu

In January of 1616, the month before before the Roman Inquisition would infamously condemn the Copernican theory as being "foolish and absurd in philosophy", Monsignor Francesco Ingoli addressed Galileo Galilei with an essay entitled "Disputation concerning the location and rest of Earth against the system of Copernicus".  A rendition of this essay into English, along with the full text of the essay in the original Latin, is provided in this paper.  The essay, upon which the Inquisition condemnation was likely based, lists mathematical, physical, and theological arguments against the Copernican theory.  Ingoli asks Galileo to respond to those mathematical and physical arguments that are "more weighty", and does not ask him to respond to the theological arguments at all.  The mathematical and physical arguments Ingoli presents are largely the anti-Copernican arguments of the great Danish astronomer Tycho Brahe; one of these (an argument based on measurements of the apparent sizes of stars) was all but unanswerable.  Ingoli's emphasis on the scientific arguments of Brahe, and his lack of emphasis on theological arguments, raises the question of whether the condemnation of the Copernican theory was, in contrast to how it is usually viewed, essentially scientific in nature, following the ideas of Brahe.

Page 2 of 60Page 2 of 60

> "All have said, the said proposition to be foolish and absurd in philosophy; and formally heretical, since it expressly contradicts the meaning of sacred Scripture in many places according to the quality of the words and according to the common explanation and sense of the Holy Fathers and the doctors of theology."[1]
>
> -- statement regarding the Copernican theory from a committee of eleven consultants for the Roman Inquisition, 24 February 1616.

In January of 1616 Monsignor Francesco Ingoli (1578-1649) addressed Galileo Galilei (1564-1642) with an essay entitled "Disputation concerning the location and rest of Earth against the system of Copernicus".  Ingoli was a well-connected clergyman who had engaged Galileo in debates at the home of Lorenzo Magalotti (1584-1637).  After one particular oral debate, the two agreed to put their arguments into writing, with Ingoli doing so first and Galileo responding.  Thus Ingoli produced the essay.  In all likelihood, Ingoli had been commissioned by the Inquisition to write an expert opinion on the Copernican controversy, and the essay would provide the chief direct basis for the consultants' condemnation the following

---

[1] "Omnes dixerunt, dictam propositionem esse stultam et absurdam in philosophia; et formaliter haereticam, quatenus contradicit expresse sententiis sacrae Scripturae in multis locis secundum proprietatem verborum et secundum communem expositionem et sensum Sanctorum Patrum et theologorum doctorum." -- Latin text from Favaro 1890-1909, Vol. XIX, 321, but with a semicolon between "philosophia" and "et formaliter" as is found in the original text in the Vatican manuscripts, rather than the comma as in Favaro, as noted in Finocchiaro 1989, 344 note 35.

month.² In a reply to the essay that Galileo wrote in 1624, he notes that Ingoli's arguments "were not lightly regarded by persons of authority who may have spurred the rejection of the Copernican opinion".³

Ingoli's essay was not published until it was included in Antonio Favaro's monumental *Opere di Galileo* at the end of the nineteenth century.  Since then it has been available from larger libraries.  The *Opere* is now available on-line, and thus Ingoli's essay is easily accessible.  Nonetheless, translations of it are scarce.  Here I provide a rendition of it into English, with commentary and with notes on Galileo's reply, in Appendix A.  The English rendition is somewhat abridged, as explained in the appendix, but within Appendix B is Ingoli's entire essay in the original Latin.

The essay provides a fascinating glimpse, at the eve of the condemnation of the Copernican theory, both of the reasoning of an anti-Copernican clergyman connected with the Inquisition, and of the tone with which that reasoning was communicated to Galileo.  The essay may not be what the reader would expect, especially if the reader is familiar with Galileo's 1624 reply, which is both lengthier and more combative in tone than Ingoli's essay.  Ingoli presents Galileo with a variety of arguments, which he describes as mathematical, physical, and theological.  Two particularly interesting features of the essay stand forth.

The first of these is that the essay frequently references the work of Tycho Brahe (1546-1601), the great Danish astronomer; often it cites Brahe's book *Astronomical Letters* by page number.  Brahe, a Danish nobleman of the generation between Copernicus and Galileo, was widely recognized as the finest astronomer of his era.  The Danish king had provided Brahe with his own island, and there Brahe built the best observatory and hired the most skilled staff money could buy (Figure 1).  Whereas Copernicus and Galileo worked as individuals (Copernicus did astronomy on the side while working as a canon at the Frombork cathedral, for example), Brahe ran a massive research program with a budget which, for its day, has been likened to that of NASA.[4]  Brahe's observatory was the best there was -- like a "Hubble Space Telescope" of his time.  His observational data were unsurpassed in quality and quantity.  The historian of science Owen Gingerich often illustrates Brahe's incomparable contribution to the astronomy of his time by means of the 1666 *Historia Coelestis* of Albertus Curtius, a catalog of astronomical observations made from antiquity to 1630.  The overwhelming majority of this thick book consists of data from Brahe, with all observations from antiquity to Brahe, and all observations after Brahe to 1630 (which would include Galileo's time), being only small sections coming before and after the Brahe material.  Gingerich argues that "only twice in the history of astronomy has there been such an enormous flood of new data that just changed the scenes" -- the flood from Tycho Brahe and the flood from the today's digital revolution.[5]  Gingerich notes that Brahe's quest for better observational accuracy "places him far more securely in the mainstream of modern astronomy than Copernicus himself".[6]

---
2  Fantoli 1994, 240-241; Finocchiaro 2010, 72; Finocchiaro 1989, 347 note 2.
3  Galilei 1624, 155.

    To Tycho Brahe, the Copernican theory had merit but also significant flaws.  Two centuries before Isaac Newton, the accepted geocentric model for the workings of the heavens was that of the ancient Greek astronomer Ptolemy, and the accepted physics was that of Aristotle.  In Aristotle's physics, heavy materials such as comprise

---

4  Couper, Henbest & Clarke 2007, 120; Thoren 1990, 188.
5  Gingerich 2009, 10:00 mark and following.
6  Gingerich 1973, 87.

objects on Earth were believed to have a natural state of rest; a wagon loaded with rock comes promptly to rest unless forced to move.  Heavenly bodies, by contrast, were believed to be comprised of a light and naturally-moving aethereal substance not found on Earth; in contrast to objects on Earth, celestial bodies keep moving eternally, ensuring that the Sun would always rise tomorrow.  The Copernican system's rapidly moving Earth was contrary to this accepted physics.  Thus Brahe admired the Copernican system as a matter of mathematics, but he objected to its assigning rapid, complex motion to the Earth, "that hulking, lazy body, unfit for motion".[7]

Yet Brahe's opposition to the Copernican system was not merely a matter of adherence to Aristotelian ideas.  Brahe produced an anti-Copernican argument that was based on his measurements of the apparent sizes of stars.  It was quite robust, scientifically speaking; it would not be fully answered until the early nineteenth century.  It was robust because it rested not on any particular physics, but only on observation, measurement, and geometry.  It said that calculations of star sizes indicated that, if the Copernican theory were correct, every last star would be enormous.  Even the smallest would dwarf the Sun like a basketball dwarfs a BB.  The Sun and its system would be tiny and unique in a universe of titanic bodies.  Brahe thought this absurd.  Ingoli calls particular attention to this argument.[8]

Ingoli also cites a second robust anti-Copernican argument of Brahe's.  This said that any rotation Earth might have ought to be detectable in the motions of projectiles fired at right angles to each other.  The Italian Jesuit astronomer Giovanni Battista Riccioli would further develop and strengthen this argument, which was also a matter

---

7  Gingerich 1993, 181.

8  Remarkably, Ingoli even suggests a solution, one that Galileo later overlooked or misunderstood in his 1624 reply to the essay (see Appendix A).

of geometry.  (A more detailed and technical discussion of both arguments is included in Appendix A.)

Brahe proposed an alternative to the Copernican theory, one which was not vulnerable to these arguments.  Brahe's alternative retained the Copernican theory's mathematical elegance, while not ascribing motion to a hulking, lazy Earth.  In this "Tychonic" theory the Sun, Moon, and stars circled an immobile Earth, while the planets circled the Sun (Figure 2).  This theory was mathematically identical to that of Copernicus insofar as the planets were concerned.  Thus, like the Copernican theory, it was compatible with Galileo's telescopic discoveries (Venus circled the Sun in both, for example).  But the Tychonic theory's fixed Earth allowed for a much smaller universe in which stars could be relatively close by and of reasonable size (Figure 3).  It did not create the expectation that evidence for Earth's motion could be detected in the trajectories of cannon balls and the like.  Furthermore, it did not conflict with scriptural passages that spoke of an immobile Earth or a moving Sun.  It was, Brahe said, a theory that "...offended neither the principles of physics nor Holy Scripture."[9]

The other particularly interesting feature of the essay that stands forth is its closing paragraph.  Here Ingoli suggests that Galileo respond to the "more weighty" of the mathematical and physical arguments, rather than to the theological arguments against the Copernican theory, or to the lesser mathematical and physical arguments.

Indeed, most of the essay is mathematical and physical arguments. The emphasis on matters of science rather than theology seems surprising, but regarding one of the theological arguments Ingoli notes that, "in explaining Sacred Writings the rule is to always save the literal sense, when it can be done, as in our case [via the Tychonic system]".  Cardinal Robert Bellarmine had emphasized this idea some months earlier in April 1615, stating in a letter that if solid evidence for the Copernican theory were found then the literal sense of scripture would have to give way to a different interpretation, but not until then.[10]

---

9  Gingerich & Voelkel 1998, 1.

This view seems to have been common in Jesuit circles, but the French Jesuit Honore Fabri (1607-1688) would be the first to publish it, writing in 1661 that --

> ...nothing hinders that the Church may understand those Scriptural passages that speak of this matter in a literal sense, and declare that they should be so understood as long as the contrary is not evinced by any demonstration....[11]

-- and continuing on to say that if some demonstration of the Copernican theory's validity were found, the Church would not scruple to declare that those passages are to be understood in a figurative sense.[12]  As solid evidence for the Copernican theory began to be found, and Rome began to reconsider the prohibitions against the theory, the Jesuit and Inquisition consultant Pietro Lazzari argued in 1757 for lifting such prohibitions.  Lazzari said that the evidence was indeed against Copernicus in 1616, and that the theory "was rejected and branded with serious objections by most excellent astronomers and physicists".  But, he said, the development of Isaac Newton's physics (which could explain Earth's motion and a variety of other phenomena in a Copernican universe, but not in a Tychonic one) in the late seventeenth century, and the detection of an effect of Earth's motion on starlight in 1728 by the English astronomer James Bradley (an effect known as *stellar aberration*), meant the literal sense could no longer be saved.[13]  The Dominican friar Maurizio Benedetto Olivieri (1769–1845), another Inquisition consultant, would put forth the same arguments in the 1820's as the Copernican theory was again being discussed in the face of still further discoveries.[14]

---

10 Bellarmine 1615, 68; Graney 2011 (a), 70-72.
11 Finocchiaro 2005, 93.
12 Finocchiaro 2005, 93-94.

Thus the mathematical and philosophical arguments against the Copernican theory would be of great importance.  If they could not be answered and Tycho Brahe's ideas still stood, theological arguments could come into play.

Ingoli's reliance on Brahe's ideas, and his suggestion that Galileo focus on the more weighty mathematical and physical arguments he presents, rather than on theological arguments, raises a question.  Opposition to the Copernican theory is often depicted as a matter of adherence to Aristotle or to religion.  Albert Einstein famously characterized such opposition as a product of "anthropocentric and mythical thinking" rooted in the "rigid authoritarian tradition of the Dark Ages" and as "opinions which had no basis but authority".[15]  Yet Ingoli's essay, written a matter of weeks before the condemnation of the Copernican theory by the Inquisition's consultants, and likely the basis of that condemnation, rests neither on references to religion nor on references to Aristotle.  Instead, it references the anti-Copernican arguments of the preeminent astronomer of the time, some of which were weighty indeed.

---

13 Finocchiaro 2005, 147-148.
14 Finocchiaro 2005, 219.
15 Galilei 2001, xxiii, xxviii.

The question then, is whether the Inquisition's condemnation of the Copernican system as "foolish and absurd in philosophy", or in more modern language, as "philosophically and scientifically untenable"[16], was to a certain extent a judgment motivated by science, rather than just by ossified intellectual tradition or by religion. The Inquisitions' "foolish and absurd in philosophy" statement is then followed by the assessment that the theory is also "formally heretical, since it expressly contradicts the meaning of sacred Scripture". According to the expert in the field, Brahe, science does not support the Copernican theory. Indeed, he believes it to be absurd. Thus the literal sense of Scripture can be saved, theological arguments come into play, and the Copernican theory, in contradicting the literal sense, is "heretical".

Brahe's scientifically robust anti-Copernican arguments would stand until astronomers came to realize that the stars did not operate in the same way as other things seen and measured by telescopic or non-telescopic instruments, or until they detected Earth's rotation through experiment. These things would be decades in coming, and would take until the nineteenth century to be fully worked out. But as long as Brahe's arguments stood, the Copernican theory, especially with its giant stars which some Copernicans explained by way of Divine Omnipotence, would be vulnerable to the charge of being an absurd theory -- vulnerable not from a scriptural standpoint, but from a scientific one. Such was Tycho Brahe's assessment, and also, following Brahe, the Inquisition's.

Monsignor Francesco Ingoli's essay to Galileo highlights the role of science in anti-Copernican thought on the eve of the Inquisition's condemnation of the Copernican theory. The essay seems not to be written by a man locked into anthropocentric and mythical ideas, rigid authoritarian tradition, or opinions with no basis but authority, but by a man informed of and interested in weighty scientific argument taken from the work of a leading scientist.  Indeed, several years after the 1616 condemnation, Ingoli asked that the Inquisition move forward to produce an approved version of Copernicus's *On the Revolutions*, as it was "most useful and necessary to astronomy".[17]  The Tychonic theory and the Copernican theory were identical from a mathematical and astronomical perspective insofar as the sun, Moon, and planets were concerned.  Everything Copernicus did was fully applicable in a Tychonic universe -- it just needed to be recognized as being a matter of convenient calculation rather than a true description of the universe.  Thus, in 1620, a revised version of Copernicus's book was produced.  Ten edits had been made to remove or change language that described the Earth as actually moving, rather than hypothetically moving.  Copernicus was thus brought into step with the universe of the great Tycho Brahe, thanks to Monsignor Francesco Ingoli, who seems to defy the usual depiction of an anti-Copernican clergyman, and whose work suggests that Inquisition opposition to Copernicus was motivated to a certain extent by scientific considerations.

---

16 Finocchiaro 1989, 29.

---

17 Tutino 2010, 277.

# APPENDIX A:  A RENDITION INTO ENGLISH OF FRANCESCO INGOLI'S 1616 ESSAY, WITH COMMENTARY AND NOTES ON GALILEO'S 1624 REPLY TO THE ESSAY

What follows is an effort to render Ingoli's Latin into a form relatively accessible to the modern reader.  While we (I thank my wife, Christina Graney, for her invaluable assistance in translating the work of Ingoli and Tycho Brahe) strove to produce a faithful rendering of Ingoli's work, there were places where we felt the need to break the essay into additional paragraphs, or to paraphrase somewhat, or to omit technical details that distract from the flow of the essay (especially if they made reference to incorrect measurements commonly accepted at the time -- such as the distance from the Earth to the Sun being roughly 1200 Earth semidiameters, for example).  Our paragraph breaks are indicated by indentations; Ingoli's original paragraphs are indicated by spacing.  Omitted material is indicated with ellipses and brackets.  For the reader who wishes to see the essay as Ingoli wrote it, Ingoli's original Latin, as included in Favaro's *Opere*, appears in Appendix B.

    Included with this rendition are comments on aspects of the essay, and notes on Galileo's 1624 reply.  Despite Galileo ignoring Ingoli's theological arguments (note Ingoli's last paragraph), Galileo's reply, at over 20,000 words, dwarfs Ingoli's essay, whose length is approximately 3000 words.  Thus the notes are limited to brief highlights toward providing a sense of how Galileo responded to the arguments.

*Francesco Ingoli*

*Ravenna*

*Disputation concerning the location and rest of Earth against the Copernican system*

*to the learned Florentine mathematician*

*D. Galileo Galilei*

*public professor of mathematics, formerly in the gymnasia to the Paduans, but now the philosopher and chief mathematician of the Most Serene Grand Duke of Etruria.*

*PREFACE*

*Among the many disputations which have come before the Very Distinguished and Most Reverend D. Lorenzo Magalotti (a committed man of prudence and letters in the Roman Curia), a particular and singular one has been that concerning the situation and motion of the Earth, as in the position of Copernicus. In this disputation you, learned man, certainly are the defender of Copernicus, offering arguments to solve those of Ptolemy, and endeavoring to confirm the Copernican system. But I have been given the role of defending the other side, bringing forth arguments to sustain the hypothesis of the old mathematicians, and to tear down the Copernican assumption. Now finally, things have come to the point for the truth to be demonstrated about your promised*

*experimental solution to the argument of Ptolemy, and for the argument proposed by me concerning parallax to be presented in writing, in order that you might be able to produce a timely solution of it.*

*I have agreed to my role extremely willingly. I am always thankful to be among the most learned men, such as yourself, engaged in the best sort of debate. Indeed I am honored by such debates, and in them I often learn some things.*

*And so now, back home, I need to fulfill my duties. But in thinking of this it occurred to me to contact you, because you might willingly hear each and every argument in this disputation which might adduce reasoning against Copernicus. Thus in order that the truth of the thing might be investigated more easily, I have resolved to write not only the argument concerning parallax, but others likewise (although not all), which can be made against the Copernican system and the motions of the Earth devised from it. If you too would think it worthwhile to write to satisfy these arguments, it will be most pleasing to me, and I will be most grateful to you.*

CHAPTER ONE: THE STRUCTURE OF THIS LETTER

I will follow this method in this disputation: I shall discuss first against the positions of the Earth and the Sun, specified by Copernicus in his system, and second against the motions of the Earth and the immobility of the Sun. I will generate three types of arguments: mathematical, physical, and theological.

*CHAPTER TWO:  MATHEMATICAL ARGUMENTS AGAINST THE COPERNICAN POSITION OF THE EARTH*

*Copernicus proposes the Sun to be in the center of the universe, while the Earth is in a circle between the orbits of Venus and Mars.*

*Against this, first I present the argument concerning parallax.  For if the Sun might be in the center of the universe, it might show greater parallax than the Moon.  But it does not.  Therefore, it is not....*

> This argument, which Galileo acknowledges as Ingoli's own new argument[18], is based on a certain astronomical measurement called the *diurnal parallax* (a difference in the apparent position against the stars of a celestial body as seen from different points on Earth's surface) being greater for the Moon than for the Sun.  This measurement is dependent only on triangle geometry, the size of Earth, and the distance to the body in question: the closer the object, the greater the diurnal parallax.  The Moon has greater diurnal parallax than the Sun.
> 	Ingoli proceeds to discuss how astronomers note in their theories and tables that the further removed a body is from the prime mover, the greater its parallax.  The parallax of the Sun at apogee is smaller than at perigee, because it is closer to the prime mover at apogee (when the Sun is most distant from Earth) than at perigee (when it is closest).  But following Copernicus, he says, the Sun is more distant from the prime mover than the Moon, because the Sun is in the center while the Moon is not, and the center is the point most distant from the periphery.  Thus the Sun should exhibit greater parallax.

---

18 Galilei 1624, 157.

Ingoli says that clearly is not the case. He cites G. A. Magini (Galileo notes that Magini treats the question of parallax extensively[19]) for measurements of that time showing the parallax of the Moon to surpass that of the Sun by 22:1. Ingoli proceeds to cite additional numbers on the distances of the Sun and Moon to bolster this argument.

However, as Galileo notes in his reply to Ingoli[20], Ingoli falsely attributes greater diurnal parallax to greater distance from the starry orb, rather than to lesser distance from Earth, ignoring that within a geocentric cosmology, in which Earth is located at the center of the orb of the stars, closer to the Earth implies further from the starry orb. Ingoli then declares that in the Copernican system, since the Sun is a greater distance from the starry orb than the Moon, it should have the greater parallax; and since it does not, Copernicus is wrong.

Basic geometry reveals the speciousness of this argument. This argument is weak in comparison to the ones that follow; one wonders whether Ingoli was either trying to throw Galileo an easy first pitch, or dispensing with a poor argument that he felt obliged to include for unstated reasons.

*The second argument is from Sacrobosco* [John of Holywood]*, in* The Sphere*, chapter 6. This says Earth is in the center of the eighth orb [the orb of the stars], because the stars appear to us to have the same size regardless of their elevation above the horizon. This would not be true if Earth did not occupy the center. This is proven from the definition of the circle (only lines which lead from the center to the circumference are equal to each other) and from the rule of view (those things which appear larger to us are closer, because they*

---

19 Galilei 1624, 162.
20 Galilei 1624, 162-5.

*subtend a greater angle; those which appear smaller are more remote, because they subtend a smaller angle).*

*The third argument is from Ptolemy, book 1, chapter 5, of the* Almagest. *This says that the Earth is at the center of the universe because an observer always sees half of the celestial sphere....*

> Ingoli also notes that the angles between certain stars, such as Aldebaran and Antares, or between other points on the celestial sphere, do not change. This would not be true if Earth did not occupy the center.
>   These arguments address the lack any changes in the appearance of the stars owing to Earth's being in an orbit about the Sun, a phenomenon called *annual parallax*. The Copernican solution to both is that the orbit of the Earth is negligibly small relative to the distances to the stars. Such effects must exist. After all, if Earth is in an orbit about the Sun, then at any given time it is, as Ingoli notes, not equidistant from all stars. Being closer to or farther from certain stars at certain times should affect their apparent size or brightness, and our perspective on them, much as Sacrobosco and Ptolemy say. But Copernicus argued that the stars are so distant that the effects are negligibly small. As he wrote in his 1543 *On The Revolutions of Heavenly Spheres*,
>
>> But that there are no such appearances [of annual parallax] among the fixed stars argues that they are at an immense height away, which makes the circle of annual movement or its image disappear from before our eyes since every visible thing has a certain distance beyond which it is no longer seen, as is shown in optics. For the brilliance of their lights shows that there is a very great distance between Saturn the highest of the planets and the sphere of the fixed stars. It is by this mark in particular that they are distinguished from the planets, as it is proper to have the greatest difference between the moved and the unmoved. How exceedingly fine is the godlike work of the Best and Greatest Artist![21]

Galileo notes this Copernican solution[22] in replying to Ingoli, but Ingoli has an answer to it. He puts forth this answer immediately following the third argument --

*Nor satisfies entirely the solution by which is said that the diameter of the circle of the deferent [orbit] of Earth in comparison to the large distance of the eighth orb [the stars] from us is so small [as to yield an effect too small to measure]. For as Tycho Brahe says in his book of Astronomical Letters, responding to Rothmann (page 188): for the Earth to be of insensible size in comparison to the starry orb, it is necessary that Earth be distant from the starry orb by fourteen thousand of its own semidiameters. And so in the Copernican system, for the Earth's orbit ... to be of insensible size in comparison to the starry orb, it must be distant by 14,000 of its own semidiameters [i.e. the stars must be at least 14,000 times farther away than the Sun]. This great distance shows the universe to be asymmetrical. But it also clearly proves ... the fixed stars to be of such size, as they may surpass or equal the size of the deferent circle of the Earth itself.... This can be proven from the apparent size of the body of the Sun; for if the Sun is seen by us to have a diameter 32' [32/60 of a degree] at a distance from the Earth of [1 solar distance], how great ought to be the size of the fixed stars,*

---

21 Copernicus 1995, 26-27.
22 Galilei 1624, 166.

*which are distant from Earth by [14,000 solar distances], as they may appear to us to be 3', following the old opinion, or indeed 2', if you prefer? Indeed by reason of these I think, the arguments of Sacrobosco and Ptolemy to be able to be solved insufficiently through the assumption that the diameter of the deferent of the Earth is [negligibly small] compared to the firmament of the sky.*

While Ingoli does not call this a separate argument, here perhaps is the weightiest part of his essay. As he notes, this is Tycho Brahe's argument concerning the sizes of stars in the Copernican theory. Stars appear to the naked eye to have measurable sizes. Brahe had measured the more prominent stars to be about one fifteenth the diameter of the Sun or Moon, or 2/60 of a degree in size (2'), and lesser stars to be progressively smaller; his values were similar to earlier measurements by Ptolemy.[23] The problem for Copernicans was that the more distant an object of a given apparent size is, the larger it must be: the Sun and Moon have roughly the same apparent size, as can be seen on an evening when the Sun is setting while the Moon is rising; but the Sun, being the much more distant, is the much larger body. By the same geometry, the more distant the stars had to be to explain the lack of annual parallax in the Copernican system, the larger they had to be. According to Brahe's measurements and calculations, an average star would be as large as the orbit of the Earth, utterly dwarfing everything in the solar system, even the Sun, whereas in a geocentric cosmos the stars lay just beyond Saturn (Figure 2), and were of reasonable size, consistent with the other celestial bodies (Figure 3).

Here was a weighty argument. As Albert van Helden has put it "Tycho's logic was impeccable; his measurements above reproach. A Copernican simply had to accept the results of this argument."[24]

---

23 Van Helden 1985, 27, 30, 32, 50.
24 Van Helden 1985, 51. If Van Helden's statement seems startling, note that Brahe's argument was once better known. An old encyclopedia article ("Brahé, Tycho" 1836, 326) includes,

Against this argument Copernicans could only appeal to God. For example, the Copernican Christoph Rothmann, when Brahe leveled this argument against him, responded,

> ...what is so absurd about [an average star] having size equal to the whole [orbit of Earth]? What of this is contrary to divine will, or is impossible by divine Nature, or is inadmissible by infinite Nature? These things must be entirely demonstrated by you, if you will wish to infer from here anything of the absurd. These things which vulgar sorts see as absurd at first glance are not easily charged with absurdity, for in fact divine Sapience and Majesty is far greater than they understand. Grant the Vastness of the Universe and the Sizes of the stars to be as great as you like -- these will still bear no proportion to the infinite Creator. It reckons that the greater the King, so much more

---

The stars, to the naked eye, present diameters varying from a quarter of a minute of space, or less, to as much as two minutes. The telescope was not then invented which shows that this is an optical delusion, and that they are points of immeasurably small diameter. It was certain to Tycho Brahe, that if the earth did move, the whole motion of the earth in its orbit did not alter the place of the stars by two minutes, and that consequently they must be so distant, that to have two minutes of apparent diameter, they must be spheres as great a radius at least as the distance from the sun to the earth. This latter distance Tycho Brahe supposed to be 1150 times the semi-diameter of the earth, and the sun about 180 times as great [i.e. by volume] as the earth. Both suppositions are grossly incorrect; but they were common ground, being nearly those of Ptolemy and Copernicus. It followed then, for any thing a real *Copernican* could show to the contrary, that some of the fixed stars must be 1520 millions of times as great as the earth, or nine millions of times as great as they supposed the sun to be.... The stars were spheres of visible magnitude, and are so still; nobody can deny it who looks at the heavens without a telescope; did Tycho reason wrong because he did not know a fact which could only be known by an instrument invented after his death?

> greater and larger the palace befitting his Majesty. So how great a palace do you reckon is fitting to GOD?[25]

Brahe viewed Rothmann's response as absurd —

> On what is such an assertion based? Where in nature do we see the Will of God acting in an irregular or disorderly manner? In nature where all things are well-ordered in all ways of time, measure, and weight? In nature where there is nothing empty, nothing irrational, nothing disproportionate or inharmonious. Consider [the vast distance] between Saturn and the fixed stars in the Copernican hypothesis. Consider those same fixed stars each being as large as the whole Orbit of Earth (and some larger still) and thus dwarfing the Sun, the luminary and center of motion for all the planets. These are the same fixed stars which are noted as the least of the heavenly lights in the account of the Creation of the World. This is empty, irrational, disproportionate and inharmonious. Is such a disproportionate universe reasonable?[26]

---

25 Graney 2012 (a), 217. For more on Copernican reactions to this problem, see Graney 2012 (b), 100-110; and Graney 2013.

26 "Ad quid hoc est dicere? Num uoluntas Diuina irregulariter et in ordinate uspiam agit, contra, quam alias in toto Mundi Theatro apparet? Vbi omnia iusto tempore, mensura et pondere undiquaque rite disposita sunt? Nihil uacui, nihil frustranei, nihil sibi inuicem non certa harmonia, et proportione correspondens. Scilicet a Saturno ad fixas Stellas non quidpiam in usus Terricolarum destinatum continebitur, per interuallum, plusquam 700000. Semidiametrorum Terrae, Stellis fixis quae longe superius elatae sunt, his tamen non minimum inseruientibus. Scilicet etiam fixa sidera nonnulla totum Orbem Annuum, quem Sol describit (siue, ut ille uult, Terra) sua magnitudine aequabunt, nonnulla uero adhuc longe maiora erunt. Ipse vero Sol, praecipuum Mundi corpus ac luminare huic quantitati uix conferendus uideatur, cum tamen Stellae in Coeli expanso constitutae, minimam, respectu Solis in patefactione Creationis Mundi obtineant praerogatiuam, uti etiam per se comparatione eius quam minimam habent; Sed quasi ab Authoritate huius, et praeminentia, uti et reliqui Planetae dependeant. Et qualis quaeso Mundi uisibilis Symmetria sic prodibit,

Brahe goes on to remark that by Divine providence such an ageometrical, asymmetrical, disorderly, and most unworthy method of philosophizing will go away.[27]  Indeed Rothmann would give up his Copernicanism and adopt Brahe's version of geocentrism.[28]

In his 1624 reply to Ingoli, Galileo answered that the telescope showed stars to measure much smaller in diameter than Brahe had measured with non-telescopic instruments.[29]  But Galileo's argument was not valid.  The telescope did produce much smaller measurements of star diameters, which would reduce the true size of the stars.  But it also increased sensitivity to annual parallax by a factor similar to that by which it reduced star diameters.  This required the stars to be even more distant, which would increase the true size of the stars.  The net result was that the stars still had to be absurdly huge.  Brahe's argument still stood.  Simon Marius argued in his 1614 *The Jovian World* that stars seen through a telescope showed themselves to be disks, and that these disks supported Brahe's geo-heliocentric theory.[30]  Giovanni Battista Riccioli would later add further support to the idea that telescopic observations of stars supported rather than undermined Brahe's argument.[31]

Remarkably, Ingoli actually suggests to Galileo the solution to Brahe's argument.  Ingoli says that the great distance either proves the fixed stars to be of enormous size, or, he says (at the point of the ellipses between "clearly proves" and "the fixed stars" in the English rendition), it proves that the fixed stars do not

---

  si maxima pars Creaturis uisibilibus destituta erit:  Quaedam uero corpora coelestia in immensum ferme augeantur, quaedam uero utut per se uasta, cum his tamen uix conferri possint?"  Brahe 1601, 191-192.

27 "Facessat ista ageometrica, et asymmetra, inordinataque philosophandi ratio, a sapientia, prouidentiaque Diuina alienissima...."  Brahe 1601, 192.

28 Barker 2004.

29 Galilei 1624, 167, 174.

30 Graney 2010 (a).

31 Graney 2010 (b).

function like other celestial bodies seen from Earth, on account of their great distance.  He offers, as an analogy, that the Sun has less effect the farther it is from the zenith (in winter versus in summer).  In his 1624 reply, Galileo interprets this statement as Ingoli saying that the stars cannot be distant because it would destroy their ability to affect things on Earth (perhaps in an astrological sense)[32], but Ingoli is clearly speaking in regards to the apparent sizes of stars (see Appendix B):

> quae distantia adeo magna non solum asymmetrum esse universum ostendit, sed etiam convincit, aut stellas fixas nihil operari posse in haec inferiora, propter nimiam earum distantiam[33]; aut stellas fixas tantae magnitudinis esse, ut superent aut aequent magnitudinem ipsius circuli deferentis Terram

which, translated closely, reads

> such a truly great distance not only reveals the universe to be asymmetric, but also clearly proves, either the fixed stars to be unable to operate in these lower regions, on account of the excessive distance of them; or the fixed stars to be of such size, as to surpass or equal the size of the deferent circle of Earth itself.

A loose translation might read, "either the stars must be huge, or something doesn't work like we think it does".
    Indeed, something did not work as as they thought it did -- both eye and telescope turn out to be unreliable for measuring the sizes of stars.  Today we know that stars are, in fact, immensely

---

32 Galilei 1624, 172.
33 Ingoli's parenthetical comment about the Sun, located here before the semicolon, is omitted here.

distant.  That immense distance means stars behave as mathematically infinitesimal points of light, whose interaction with the optics of the eye or the telescope produce, via diffraction and other effects, a spurious appearance of size (Figure 4), even when seen through early telescopes.  Only decades after Galileo's death would astronomers begin to grasp this, and a full understanding of the spurious nature of the apparent sizes of stars seen through a telescope would not be developed until the early nineteenth century.[34]

*The fourth argument is from Tycho, in the Astronomical Letters, page 209, where he proves by most certain experiments...*

From here Ingoli devotes roughly fifty words to describe how, as Galileo would put it, "the eccentricities of Venus and Mars are different from what Copernicus assumed, and likewise the apogee of Venus is not stable, as he believed".[35]  These are minor details which can be corrected, Galileo notes, and they are "quantities that have nothing to do with the stability or location of the sun or the earth".[36]

CHAPTER 3:   PHYSICAL ARGUMENTS

*Two arguments seem clear to me for Earth to be in the center of the universe.  One is taken from the order of the universe itself.  For we see in the arrangement of simple bodies, the denser and heavier to occupy the lower place, as is well known concerning earth with respect to water and concerning water with respect to air.  Earth is denser*

---

[34] Graney & Grayson 2011.
[35] Galilei 1624, 174.
[36] Galilei 1624, 175.

*and heavier than the solar body, and the lower place in the universe is without doubt the center. Thus the Earth, and not the Sun, occupies the middle or center of the universe.*

*But the first part of the minor proposition of this argument can be proven if it were to be denied. First, by the authority of the Philosopher [Aristotle] and all the Peripatetics [Aristotelians], saying the heavenly bodies to have no heaviness. Second, by logical reasoning at least, for the opposite proposition is this: the Sun is a body denser and heavier than Earth. At first thought this is seen to be false, since all things which have luminosity we may see to be rarer and lighter, as is well known concerning fire and that which is thrown out from fire.*

*Truly the second part can be proven if it were to be denied, by the authorities of the philosophers, saying the position of the center of the universe to be a place down, or lower, and the circumference of the same to be a place up, or higher. And it can be proven by reasoning, because in the globe of the Earth itself we designate as higher, parts which are located near the periphery of it, and lower those toward the center. Thus we may say the lowest part of Earth to be the center itself. The center therefore is the lower place in the universe.*

*The second argument is one taken from bits of the Earth itself. For in wheat needing to be sifted we see lumps of dirt, which are in the wheat itself, restored by the circular motion of the sieve to the center of the sieve itself. The same happens in the case of heavier bits of gravel, while they are stirred in any round vessel. By such*

*experiments many philosophers have volunteered that the Earth stands in the middle of the universe, because that way it is dispossessed from the motions of the heavens.  If this happens to bits of the Earth, likewise it must be said to happen to the whole of Earth, seeing that it may represent the argument from part to whole in the homogeneous.  Rothmann in his letter, which is in the book of Astronomical Letters of Tycho, page 185, makes use of this genus of argument from parts of Earth to the whole Earth in defending Copernicus.*

> Galileo spends a couple thousand words answering the roughly 500 words of these two arguments.[37]  To what is perhaps the strongest part of Ingoli's arguments -- the argument that fire and other luminous material is light and rises upward, and thus the Sun above would be expected to be of a light material -- Galileo responds with an ingenious counter-argument:  Earthly fire is brief in duration because it is a rarefied substance; since the Sun burns eternally, that shows it to be an extremely dense and solid substance.[38]

CHAPTER 4:   THEOLOGICAL ARGUMENTS

*Finally, as I shall conclude the first part of this disputation that the Earth and not the Sun is in the middle of the Universe, two other arguments, taken from Sacred Scripture and from the doctrine of the theologians, stand out to me.  One of these is taken from chapter 1 of Genesis.  Ponder the words: "And God said: Let there be lights made in the firmament of heaven."[39]  For in the Hebrew text the name* רקיע *rakia*

---

37 Galilei 1624, 175-181.
38 Galilei 1624, 180.
39 Genesis 1:14, Douay-Rheims translation.

*signifies expanse or extent or extension; one of these words may be considered in the place of the word* firmament, *as Saint Pagninus recommends in the Thesaurus of the Holy Language in regard to the root* raka. *And a meaning of this sort is in no way appropriate to a center, the nature of a center itself being repugnant to an extension or, as thus I may say, to an expansion. It may, however, be suitably appropriate to the circumference of the heaven, which is in a certain way extended and expanded above the center (whence, the appropriate metaphor in Psalm 103:2 referring to God "Who stretchest out the sky like a pavilion"[40]). The greater fact needing to be said is that God said, "Let lights be made in the firmament of heaven" -- not in the center to illuminate, but in the expanse or extent itself of heaven. This argument is strengthened because the word* Fiant [third person plural verb]*, which God has said, considers equally Sun and Moon, since in the text it is said, "Fiant luminaria in firmamento coeli"; whence as the Moon is not in the center, but in the expanse of the heaven, thus also the Sun ought to be in the expanse, and not in the center.*

*Another argument is from the doctrine of theologians. By it this holds mainly on account of the reasoning that hell, that is the place of the demons and of the damned, be in the center of Earth, because, since heaven may be the place of the angels and the blessed, it behooves the place of the demons and the damned to be in the most remote place from heaven, which is the center of Earth. Thus Hell and Heaven are appointed places most distant from each other, as Psalm 138 has said: "If I ascend into heaven, thou art there: if I descend into*

---

40 Douay-Rheims translation.  This is Psalm 104:2 in the New Revised Standard Version, in which the translation reads "You stretch out the heavens like a tent".

*hell, thou art present."*[41]  *Also Isaiah 14, where is said to the king of Babylon, and to the devil in the shape of him: "And thou saidst in thy heart: I will ascend into heaven.... But yet thou shalt be brought down to hell, into the depth of the pit."*[42]  *Read the Most Illustrious Cardinal Bellarmine,* De Christo, *4th book, 10th chapter, and* De Purgatorio, *2nd book, 6th chapter.  And so since hell is in the center of Earth, and hell ought to be the most remote place from heaven, Earth is admitted to be in the middle of the universe, which is the most remote place from heaven.  This marks the end of the first part of this disputation.*

> Galileo does not address these theological arguments.  See the last paragraph of Ingoli's essay.

CHAPTER 5:   MATHEMATICAL ARGUMENTS AGAINST THE COPERNICAN MOTION OF THE EARTH.

*Many things are able to be presented against the diurnal motion of Earth.  Some of these Tycho directs against Rothmann, the defender of the opinion of Copernicus, in two astronomical letters in the book of Astronomical Letters, pages 167 and 188.  One is concerning the fall of a lead ball perpendicularly from the highest tower ... because Earth by diurnal motion, even in the northern parallels of Germany, might be moved 150 greater paces in a small second of time.  A similar one is concerning the bombards [large cannon] discharged from east*

---

41 Psalm 138:8, Douay-Rheims translation.  This is Psalm 139:8 in the New Revised Standard Version, in which the translation reads, "If I ascend to heaven, you are there; if I make my bed in Sheol, you are there."

42 Isaiah 14:13,15, Douay-Rheims translation.

*into the west and from the north into the south, particularly concerning those discharged near the poles, where the movement of Earth is slowest. For, given a diurnal movement of Earth, evident differences might be observed, and nevertheless no differences are observed.*

> In his 1624 reply, Galileo spends many pages countering these two arguments by means of an argument from common motion. Galileo describes Brahe as being of the opinion that a rock dropped from atop the mast of a ship retains none of the ship's forward motion, and thus drops to the rear of the mast by as much as the ship moves forward during the time of fall, and so Brahe likewise expects a lead ball falling from a tower on a rotating Earth to fall well to the west of the tower as the Earth's rotation carries the tower eastward. Galileo writes: "I still frequently meet people with such a thick skull that I cannot put it into their head that, because the man on the mast keeps his arm still, the rock does not start from rest".[43]
>
> It is not clear that Brahe was of this opinion; the section of the *Astronomical Letters* that Ingoli cites rather suggests that Tycho believed the ball to share the forward motion of the ship or tower, but to then gradually lose that forward motion as it falls. Regardless, Tycho does argue that a ball dropped from a tower would fall at least somewhat to the west of the tower, and he notes, in the part of the *Astronomical Letters* cited by Ingoli, that he believes experiments to support this:
>
>> *Some suppose that a projectile that is launched upward from, and falls back to, the deck of a ship will land in the same place whether the ship is at rest or moving. On the contrary, the faster the ship moves, the more*

---

[43] Galilei 1624, 182-187; quote from 184-185.

> discrepancy is seen in the landing.  And likewise ought
> to happen in regards to the rotation of the Earth."[44]

Galileo also dismisses Brahe's cannon argument via common motion, discussing how a person shut inside the cabin of a ship will observe various phenomena to occur in the same way whether the ship is at rest or moving with constant speed.  In his reply he asks,

> Now, once you have made all these observations, and seen
> how these motions ... appear exactly the same when the
> ship moves as well as when it stands still, will you not
> abandon all doubt that the same must happen in regard to
> the terrestrial globe, as long as the air moves together
> with it?[45]

However, Brahe's cannon argument may have been more sophisticated than Galileo realized; certainly a more sophisticated version of the argument would be developed later by the Italian Jesuit astronomer Giovanni Battista Riccioli.  In the part of the *Astronomical Letters* that Ingoli cites, Brahe writes,

> Consider that near the poles any supposed diurnal motion
> drops to nothing.  There, the azimuthal direction at
> which a cannon was aimed would make no difference on the
> expected flight of the ball.  But at the equator, where
> the motion of the Earth's surface would be most rapid,
> there would be a difference between a ball being hurled
> to the East or West versus its being hurled North or

---

44 "Quod uero quidam exstimant telum e naui sursum eiectum, si intra nauis latera id fieret, casurum in eundem locum mota naui, quam pertingeret hac quiescente; inconsiderate haec proferunt, cum res longe aliter se habeat: Imo, quo uelocior erit nauis promotio, eo plus inuenietur discriminis: Pariter et in circuitu Terrae, quoad magis uel minus haec euenire oportet."  Brahe 1601, 190.

45 Galilei 1624, 187.

> South: in the first case the diurnal motion is with the ball; in the second it is not.[46]

Brahe appears to be noting that points at different latitudes on the Earth's surface move at different speeds, unlike the deck of Galileo's ship. The difference in speeds of Earth's surface does, in fact, produce an effect on projectiles that can be measured, and that does vary between the poles and the equator -- an effect known today as the *Coriolis Effect*. This is the phenomenon that lies behind the Foucault Pendulum, the common demonstration of Earth's rotation found in science museums and academic buildings -- pendulums whose rate of precession varies with latitude. In his 1651 *New Almagest*, Riccioli would further develop into a more rigorous argument the idea of a difference on a diurnally rotating Earth between the trajectories of cannonballs launched to the North or South versus launched to the East or West. He argued that the lack of any observed effect argued against Earth's motion. He would note, however, the possibility that the effect was so small as to evade detection.[47]

*Many more arguments can be presented from Tycho against Earth's annual motion, but I shall bring up only four. One is that if Earth is moved by annual motion, it ought to vary sensibly the latitudes of the fixed stars....*

---

46 "Addo uero et hoc, quod si circa Terrae polos, ubi motus diurnus (si quis esset) in quietem desinit, eadem fieret uersus quamcunque; Horizontis partem per sclopetum ratione ante dicta experimentatio, idem omnimode eueniret, ac si in medio inter utrumque; polum apud Aequatorem, ubi motio circumferentiae Terrae concitatissima esse deberet: Vti etiam in quouis Horizonte, si uersus ortum et occasum parili ratione emittatur globus, idem conficit spatij, quod uersus Meridem et Septentrionem simili impulsione emissus, cum tamen Terrae, si quis inesset diurnus motus, is occasum ortumque; respiceret: Meridiem uero et Septentrionem non item: Cum igitur haec uniformiter ubique; eueniant, quiescat etiam ubique; uniformiter Terra, necessum est." Brahe 1601, 189-190.

47 Graney 2011 (b); *Physics Today* 2011-2012.

*Another is based on the altitude angle of the pole over the northern horizon. For if Earth is moved by annual motion, the altitude of the celestial pole must change.... indeed just as a man who travels a road from south to north or vice versa, observes a change in the altitude of the pole....*

> Both of these are essentially more annual parallax arguments. Thus they are valid in that effects something like what Ingoli describes would be expected if Earth moves relative to the stars. Galileo objects to the way Ingoli describes them, stating, for example, that a moving Earth implies that
>
> > ...there would be a change, not in the elevation of the pole, but in the elevation of some fixed star, such as, for example, the nearby Polaris, and then [an anti-Copernican could] add that, since this is not seen, one could thereby infer the stability of the earth.[48]
>
> But Galileo goes on to say, "Copernicus already answered this by saying that, because of the immense distance of the fixed stars, such a change is imperceptible."[49] However, this rebuttal to annual parallax arguments brings up the star size issue (see the discussion following the "Mathematical arguments against the Copernican position of the earth").

*A third is based on the inequality of the length of days.... especially around the northernmost habitations, where the variations of the days are most sensible....*

---

48 Galilei 1624, 190.
49 Galilei 1624, 190.

> Galileo notes that in the Copernican system "the equator and its axis always keep the same inclination and direction relative to the zodiac (namely the circle of annual motion)", thus accounting for the changes in day length over a year.[50]

*A fourth is from Tycho in the book of the Astronomical Letters, page 149, where he asserts that the heavenly maneuvering of comets when opposite the Sun in the sky does not comport to an annual motion of Earth....*

> Galileo responds to this in part by criticizing Tycho, arguing that Tycho could not observe comets when in opposition to the Sun, because
>
>> ...their tail always points away from the sun, [and] then it is impossible for us to see any of them.... Furthermore, what does Tycho know with certainty about a comet's own motion, as to be able confidently to assert that, when mixed with the earth's motion, it should produce some phenomenon different from what is observed?[51]

*Finally Tycho in the Letters presented three arguments against the Earth's third motion: First, because if the annual motion is removed, the third may be necessarily taken away. Second, because it is not possible that the axis of Earth may truly gyrate correspondingly contrary to the motion of the center in an annual manner, so as to be seen to rest. Third, because it is not possible to be granted, in a single and simple body, for the axis and center to be moved by a*

---

50 Galilei 1624, 191.
51 Galilei 1624, 191.

*double and opposite motion; to which, furthermore, if the diurnal motion may be added, even more difficulty is produced.*

> The "third motion" was that by which the Earth's axis maintains the same orientation in space, parallel to itself at all times, so that it is always pointed at the North Star (Polaris).  If one builds a mechanical orrery to illustrate the Earth's motion in the Copernican system, one needs one mechanism to cause the Earth move around the Sun annually, another to cause the Earth to rotate diurnally, and a third to cause Earth's axis to remain pointed in the same direction in space at all times, rather than in the same direction with respect to the Sun, as it would be if the axis for the diurnal rotation were to be simply fixed at an angle to the arm for annual motion.  This third motion would exactly match the period of the annual motion, but be contrary to it.
>
> Today we understand that a rotating body naturally maintains its orientation in space gyroscopically, through the physics principle known as "conservation of angular momentum", but at the time this was considered an actual third motion.  Galileo responds to these arguments by citing the example of a wooden ball floating in a bowl of water: a person holding the bowl can turn around, and to that person the ball appears to turn in the water; but the ball actually maintains its orientation in space and does not truly turn with its own motion.[52]

CHAPTER 6: PHYSICAL ARGUMENTS AGAINST THE MOTION OF THE EARTH

*Many things which are produced by the philosophers and mathematicians on behalf of a resting Earth (and especially by Tycho in the Letters),*

---

52 Galilei 1624, 192.

*might adduce physical arguments against the motion of Earth; but I shall put forward only three common ones.*

*One is from the nature of heavy and light bodies. For in general we see the heavy bodies to be less apt to motion than the light or not heavy.... Since therefore Earth may be the heaviest of all bodies knowable to us, it is by no means proper to say nature to have bestowed to Earth so many motions, and especially the so-swift diurnal ... as it says in the Astronomical Letters by Tycho, page 190.*

> That the Earth was too heavy to be suitable for motion was an important point for Brahe. He had once said that the Copernican system
>
>> …expertly and completely circumvents all that is superfluous or discordant in the system of Ptolemy. On no point does it offend the principle of mathematics. Yet it ascribes to the earth, that hulking, lazy body, unfit for motion, a motion as quick as that of the aethereal torches, and a triple motion at that.[53]
>
> Galileo answers this by, among other things, denying that heavy bodies are unsuitable to motion. In fact, he says, a lead ball travels farther when fired by a cannon than a wooden one, which in turn travels farther than a wad of straw or fiber. He also notes that "I see bowl makers and tin-plate turners adding very heavy wooden wheels to their machines in order to make them retain longer the impetuses they acquire...."[54]

*Another is taken from that physical proposition which says that to a natural body there belongs only one natural motion.... Accordingly,*

---

53 Gingerich 1993, 181.
54 Galilei 1624, 193.

*since the natural motion of Earth may be towards the middle, motion around the middle will not be able to be natural to it....*

> Galileo's reply to this in 1624 was a lengthy discussion about straight and curved motion, ending with "So I conclude that if the earth has a natural inclination to motion, this can only be toward circular motion...."[55]

*The third is from a certain incongruence: because of course to all bright parts of the heaven, certainly to the planets, Copernicus has attributed motion. Yet to the Sun (of all parts of the heaven the most outstanding and bright) he denies motion, while to Earth (the dark and dense body), he assigns motion. Indeed Nature, most discrete in all of its works, ought not to do this.*

> Galileo's 1624 reply was to point out that the Sun is the source of light for the moon, planets, and Earth, none of which produce their own light. "Therefore we can very resolutely assert that the earth's conformity to the other six [bodies] is very great, and that on the contrary the discrepancy between the sun and these bodies is equally great."[56]

CHAPTER 7: THEOLOGICAL ARGUMENTS AGAINST THE MOTION OF EARTH

*Endless theological arguments from Sacred Scripture and from the authority of the Fathers and of the Scholastic theologians might be able to be proposed against the motion of Earth, but I shall adduce only two, which seem to me to be more substantial.*

---

[55] Galilei 1624, 195.
[56] Galilei 1624, 196.

*One is from Joshua, chapter 10, where on the prayers of Joshua, Scripture says: "So the Sun stood still in the midst of heaven, and hasted not to go down the space of one day. There was not before nor after so long a day, the Lord obeying the voice of a man."[57] The responses to this which are produced, namely that Scripture may speak following our manner of understanding, do not satisfy: first because in explaining Sacred Writings the rule is to always save the literal sense, when it can be done, as in our case [through the Tychonic system]; next because all the Fathers unanimously explain this passage to mean the Sun, which moved, truly stood still on the prayers of Joshua. Truly the Tridentine Synod (4th session, in the doctrine concerning the publication and use of Sacred Books, § Praeterea) is averse to any interpretation which is against the unanimous consensus of the Fathers. And granted that the Holy Synod [Council of Trent] may speak in the matter of morals and of the Holy Faith, nevertheless it is not able to be denied, but that the interpretation of the Holy Scripture against the consensus of the Fathers may displease those Holy Fathers.[58]*

*Another is by authority of the Church: for in the hymn at vespers of the third day [Tuesday] thus she sings:*

> *Earth's mighty Maker, whose command*
> *raised from the sea the solid land;*
> *and drove each billowy heap away,*
> *and bade the earth stand firm for aye.[59]*

---

57 Joshua 10:13-14, Douay-Rheims translation.

58 Ingoli is speaking of two groups of "Fathers" -- the Church Fathers (from ancient times) and the Synod Fathers of the Council of Trent.

59 Taken from Martis 2007, 632. A more close translation reads,

Page 38 of 60

*The origin of this sort of argument is not trivial. Such is seen in writings of Cardinal Bellarmine. In many passages he refutes many errors by hymns, songs and prayers of the Church, which are found in breviaries.*

> As with the earlier theological arguments, in Galileo's 1624 letter of reply he does not address these theological arguments. See the following paragraph. Note that Ingoli calls attention to Bellarmine in both sections of Theological arguments.

[CLOSING]

*These complete this disputation. Let it be your choice to respond to this either entirely or in part -- clearly at least to the mathematical and physical arguments, and then not to all of them, but to the more weighty ones. For I have written this not towards attacking your erudition and doctrine (most notable to me and to all men both inside the Roman Curia and outside), but for the investigation of the truth, which you profess yourself always to search for by all strength, and in fact thus suits a mathematical talent.*

THE END

---

> Mighty author of the Earth
> Who digging up the bottom of the World,
> Banished the troubles of the waters,
> You have given the Earth immobility.

# APPENDIX B: FRANCESCO INGOLI'S ESSAY[60]

*Francisci Ingoli*

*Ravennatis*

*De situ et quiete terrae contra Copernici systema disputatio*

*ad doctissimum mathematicum*

*D. Galilaeum Galilaeum*

*Florentinium*

*Publicum professorem mathematicarum olim in gymnasia Patavino, nunc autem philosophum et mathematicum primarium serenissimi magni ducis Etruriae.*

*PROOEMIUM*

*Inter multas disputationes quas apud Perillustrem et Reverendissimum D. Laurentium Magalottum, virum ob prudentiam et litteras in Romana Curia commendatum, habuimus, illa praecipua et singularis fuit de situ et motu Terrae iuxta positionem Coperniceam. In qua tu quidem, vir doctissime, Copernici partes defendendas assumens, plurima in medium proferebas, quibus Ptolomaei argumenta solvere, et systema Copernici comprobare, conabaris: ego autem, contra, veterum mathematicorum hypothesim sustinere, et Coperniceam assumptionem destruere, vario argumentandi genere pro viribus nitebar. Tandem, post multa, eo res devenit, ut pro solutione argumenti Ptolomaei experimento, quod pollicebaris, veritas probaretur, et argumentum de parallaxi a me propositum scripto exhiberetur, ut maturius eius solutionem afferre posses. Annui perquam libenter: nam cum viris doctissimis et in disputationibus modestissimis, qualis*

---

60 Favaro 1890-1909, Vol. V, 403-412

*tu es, agere gratissimum semper mihi fuit; aliquid enim plerunque addisco, honoremque non minimum adipiscor. Domum itaque reversus, promissa implere cogitavi: sed cum, inter cogitandum, mihi te dixisse occurrisset, quod in hac disputatione libentissime unumquemque audires, qui rationes contra Copernicum adduceret, ut sic facilius rei veritas investigaretur, deliberavi non solum de parallaxi argumentum scribere, sed alia quoque, licet non omnia, quae contra systema Coperniceum et Terrae motiones, ab eo excogitatas, fieri possunt: quibus si tu quoque scripto satisfacere dignaberis, gratissimum mihi erit, et plurimas tibi habeo gratias.*

*ORDO HUIUS SCRIPTIONIS.*

*CAP. PRIMUM*

*Methodus autem in hac disputatione a me servanda erit huiusmodi. Primo, disseram contra situationem Terrae et Solis, quam ponit Copernicus in suo systemate; 2°, contra motus terreni orbis et Solis quietem: et in utroque capite triplici argumentorum genere, videlicet mathematico, physico et theologico.*

*MATHEMATICA ARGUMENTA CONTRA SITUM TERRAE COPERNICEUM.*

*CAP. 2M.*

*Proponit Copernicus, Solem esse in centro universi, Terram autem in circulo inter Veneris et Martis orbes.*

*Contra huiusmodi positionem, primum, obiicio argumentum de parallaxi. Nam si Sol esset in centro universi, maiorem admitteret parallaxim quam Luna: sed consequens est falsum: ergo, et antecedens. Consequentia probatur: quia corpora, quanto remotiora sunt a primo mobili, in quo eorum loca notantur ab astronomis, tanto maiorem admittunt parallaxim, ut et diversitatis aspectus*

*theoricis et tabulis constat, in quibus Solis apogaei parallaxis minor notatur, quia tunc vicinior est primo mobili, maior autem perigaei, quia remotior: sed Sol, iuxta Copernicum, est remotior a primo mobili quam Luna; quia haec est extra centrum, ille vero in centro, et centrum est remotior locus a periphe­ria: igitur Sol maiorem admittet parallaxim. Falsitas vero consequentis facillime probatur: nam ex observationibus manifestum est, Solis parallaxim maiorem esse 2' 58'', Lunae vero partis 1.6' 21'', ut ex Rehinoldo annotavit Maginus, Theoricorum lib. 2°, cap. 20 in fine; ex quibus observationibus liquet, non Solis parallaxim maiorem esse parallaxi Lunae, sed hanc illam longe superare, ut nimirum numerus 22 superat unitatem. Nec satisfacit si dicatur, ideo Lunam habere maiorem parallaxim, quia nobis vicinior est, cum distet a Terra semidiametris terrenis tantum 52.17' usque ad 65.30' a quarto limite ad primum, ut ex Copernico notat Maginus, Theoricorum lib. 2°, cap. 24; quibus Sol distat 1179. Primo: quia si haec solutio valeret, necessarium esset, ut quam proportionem habent luminum distantiae inter se, eandem haberent et parallaxes eorum, cum parallaxes a distantiis pendeant: hoc autem non videmus; quia distantiae se habent sicut 18 ad 1, ut Maginus notat ex Copernico ubi supra, parallaxes autem sicut 22 ad 1, ut dictum est: igitur solutio nihil valet. Secundo: quia parallaxis quantitatem non solum efficit distantia corporum visorum in sublimi a nobis, sed etiam distantia ab octavo orbe, ubi notantur parallaxes. Cum itaque Sol distet a caelo stellato plus quam Luna, quando est in Solis opposito, iuxta observationes Copernici, semidiametris terrenis 1244, non videtur mihi fieri posse ut parallaxis Solis sit 1/22 parallaxis Lunae.*

*Secundum argumentum est Sacrobusti, in Sphaera, cap. 6, dicentis, Terram, esse in centro octavi orbis, quia stellae in quacunque elevatione sint supra horizontem, eiusdem quantitatis nobis apparent: quod non esset, si Terra centrum non possideret. Quod probatur, tum ex diffinitione circuli; nam solae lineae quae a centro ad circumferentiam ducuntur, sunt inter se aequales: tum ex regula prospectivae, qua dicitur, quae maiora nobis apparent, viciniora esse, quia sub maiori angulo videntur, quae autem minora, remotiora, quia sub minori angulo conspiciuntur.*

*Tertium argumentum est Ptolomaei, lib. 1, cap. 5, Almagesti, dicentis, ideo Terram esse in centro mundi, quia, ubicunque existat homo, semper videt coeli medietatem, hoc est gradus 180; quod non esset si Terra esset extra centrum. Quod autem coeli medietas ubicunque conspiciatur, liquet non solum ex stellis fixes oppositis, nempe ex Oculo Tauri et Corde Scorpionis, quarum una oritur dum alia, occidit; sed etiam ex certa observatione graduum 90, quae potest haberi dum Sol est in punctis Arietis vel Librae, si notetur elevatio aequatoris meridiana, et ab ea usque ad polum interiecta distantia observetur, et tandem cum hac mensuretur portio orientalis et portio occidua circuli verticalis. Quod vero medietas coeli non conspiceretur si Terra centrum non occuparet, constat ex diffinitione semicirculi: sola enim diameter, quae semper transit per circuli centrum, dividit ipsum circulum in duos semicirculos aequales. Nec solutio qua dicitur, diametrum circuli deferentis Terram in comparatione distantiae maximae octavi orbis a nobis adeo exiguam fieri, ut in ipso orbe octavo solum 20' subtendat, omnino satisfacit. Nam si Terra, ut insensibilis magnitudinis evadat respectu stellati orbis, necesse est ut distet semidiametrorum suarum quatuordecim millibus ab ipso, iuxta Tychonis placita, ut videre est in eius libro Epistolarum Astronomicarum in responsione litterarum Rothmani, pag. 188, oportebit quoque*

*ut circulus deferens Terram (cuius semidiameter est semidiametrorum terrenarum 1179, si Magino credimus, qui distantiam Solis apogei a Terra, Theoricorum lib. 2, cap. 24, tantam esse scribit iuxta Coperniceas observationes) distet ab octava sphaera suis semidiametris m/14, quae faciunt semidiametros terrenas 16506000: quae distantia adeo magna non solum asymmetrum esse universum ostendit, sed etiam convincit, aut stellas fixas nihil operari posse in haec inferiora, propter nimiam earum distantiam (quod comprobari potest ex iis quae contingunt in Sole; nam experimur virtutem eius in hyeme, propter distantiam ipsius a Zenith capitis nostri, minimam certe in comparatione distantiae Terrae ab octavo orbe, adeo hebetem fieri ut frigus magnum persentiamus); aut stellas fixas tantae magnitudinis esse, ut superent aut aequent magnitudinem ipsius circuli deferentis Terram, cuius semidiameter est, ut diximus, 1179 semidiametrorum terrenarum: quod probari potest ex magnitudine apparenti corporis solaris; nam si Sol nobis videtur diametrum habere 32' in distantia a*

*Terra semidiametrorum terrenarum 1179, quanta debebit esse magnitudo fixarum, quae distant a Terra semidiametris terrenis 16506000, ut nobis appareant esse 3', secundum antiquam opinionem, vel etiam 2', secundum placita tua? Ex his itaque puto, Sacrobusti et Ptolomaei argumenta minime solvi posse per assumptionem, quod diameter deferentis Terram subtendat solum 20' in coeli firmamento.*

*Quartum argumentum est Tychonis in dictis Epistolis Astronomicis, pag. 209, ubi probat certissimis experimentis, reperisse eccentricitates [Mars symbol] et [Venus symbol], notatas a Copernico, aliter longe se habere; sicut et apogeum Venus non esse immobile, ut idem Copernicus affirmavit, sed sub fixarum sphera moveri: ex quibus valde dubium Copernici systema efficitur, cum phaenomenis, pro quibus salvandis ab eo sic constitutum est, minime satisfaciat.*

*ARGUMENTA PHYSICA*

*CAP. 3M*

*Terram esse in medio universi, duo argumenta mihi videntur ostendere; quorum alterum est, quod ab ordine ipsius universi desumitur. Nam in coordinatione corporum simplicium videmus, crassiora gravioraque inferiorem locum occupare, ut patet de terra respectu aquae et de aqua respectu aeris: Terra autem crassius et gravius corpus est corpore solari; et locus inferior in universo procul dubio est centrum: Terra igitur, et non Sol, medium sive centrum universi tenet.*

*Quod si negetur prima pars minoris propositionis huius argumenti, potest probari, primo, authoritate Philosophi et Peripateticorum omnium, dicentium corpora caelestia nullam habere gravitatem: 2°, ratione saltem logica; nam propositio opposita, hoc est, Sol est corpus crassius et gravius Terra, ipso primo animi conceptu videtur esse falsa, cum omnia quae habent lucem videamus esse rariora et leviora, ut patet de igne et de iis quae passa sunt ab eo.*

*Si vero negetur secunda pars, et philosophorum authoritatibus probari potest, dicentium positionem centri universi esse locum deorsum, circumferentiam vero eiusdem esse locum sursum, quod est idem ac si diceretur inferius et superius: et ratione; quia in ipso Terrae globo superiores partes dicimus quae ad peripheriam eius, inferiores vero quae infra circumferentiam et versus centrum, locantur, ita ut centrum ipsum infimam dicamus esse Terrae partem. Centrum igitur est inferior locus in universo.*

*Alterum argumentum est quod a partibus ipsius Terrae desumitur. Nam in cribrando tritico videmus quod glebae terrae, quae sunt in ipso tritico, ad motionem circularem cribri ad centrum ipsius cribri reducuntur; et idem evenit in partibus sabuli crassioribus, dum aliquo rotundo in vase agitantur; quo experimento multi philosophi voluerunt, Terram in medio universi stare, quia illac a motionibus coeli detruditur: quod si partibus Terrae id contingit, toti quoque*
*Terrae id accidere dicendum est, cum in homogeneis teneat argumentum a parte ad totum, et Rothmanus in sua epistola, quae est in libro Epistolarum Astronomicarum Tychonis, pag. 185, defendens Copernicum, hoc genere argumentandi a partibus Terrae ad totam Terram utatur.*

ARGUMENTA THEOLOGICA

CAP. 4.

*Tandem, ut primam huius disputationis partem concludam, Solem non esse in medio universi, sed Terram, duo alia argumenta ex Sacris Litteris et ex doctrina theologorum desumpta, mihi ostendere videntur. Quorum alterum est ex cap. 1 Genesis, ponderando verba: Dixit Deus, Fiant luminaria in firmamento coeli. Nam, cum in textu Hebraico habeantur loco verbi* firmamento *nomen* רקיע rakia, *quod significat expansum seu extensum vel extensionem, ut probat Sanctes Pagninus in Thesauro Linguae Sanctae in radice* raka; *et huiusmodi significatio nullo modo possit convenire centro, repugnante ipsius centri natura extensioni seu, ut ita dicam, expansioni; conveniat autem aptissime coeli circumferentiae, quae*

*quodammodo est extensa et expansa supra centrum (unde, apposita metaphora, Psalmo 103-2, dicitur, Extendens (scilicet Deus) coelum sicut pellem); dum Deus dixit, Fiant luminaria in firmamento coeli, non in centro luminare maius factum esse dicendum est, sed in ipso coeli expanso seu extenso. Confirmatur haec, argumentatio ex eo, quod verbum* Fiant*, quod Deus dixit, respicit aequaliter Solem et Lunam, cum in textu dicatur, Fiant luminaria in firmamento coeli; unde sicut Luna non est in centro, sed in coeli expanso, ita quoque Sol in hoc, et non in illo, esse debet.*

*Alterum argumentum est ex doctrina theologorum, tenentium ea potissimum ratione infernum, idest locum daemonum et damnatorum, esse in centro Terrae, quia, cum coelum sit locus angelorum et beatorum, oportet locum daemonum et damnatorum esse in loco remotissimo a coelo, qui est centrum Terrae. Unde bene, Psalmo 138, apponuntur infernus et coelum tanquam loca distantissima, dum dicitur: Si ascendero in coelum, tu illic es; si descendero in infernum, ades: et, Isaiae 14, dum dicitur regi Babylonis, et in eius figura diabolo: Dixisti, In coelum conscendam, etc.; veruntamen usque ad infernum detraheris, et in profundum lacum. Legatur Illustrissimus Cardinalis Bellarminus, De Christo, lib. 4°, cap. X°, et De Purgatorio, lib. 2°, cap. 6°. Cum itaque infernus sit in centro Terrae, et debeat esse locus remotissimus a coelo, Terram esse in medio universi, qui est locus a*
*coelo remotissimus, fatendum est. Ex quibus sit impositus finis primae parti huius disputationis.*

*ARGUMENTA MATHEMATICA CONTRA MOTUM TERRAE COPERNICEUM.*

*CAP. V.*

*Contra motum Terrae diurnum multa obiici possunt, quorum aliqua contra Rothmanum, Coperniceae sententiae defensorem, in duabus epistolis astronomicis refert Tycho in libro Epistolarum Astronomicarum, pag. 167 et 188: videlicet de casu plumbei globi ab altissima turri perpendiculariter, non obstante praetensa*

*aeris concomitantia, cum tamen deberet esse contrarium, quia Terra motu diurno, etiam in parallelis borealibus Germaniae, moveretur sesquicentum passus maiores in secundo minuto temporis: item de bombardis exoneratis ab oriente in occidentem et a septentrione in austrum, praesertim de exoneratis prope polos, ubi motus Terrae tardissimus est; nam, dato motu Terrae diurno, evidentissima differentia notaretur, cum tamen nulla animadvertatur.*

*Contra vero annuum, multo plura obiici possunt, de quibus per Tychonem ubi supra; sed ego adducam tantum quatuor rationes: quarum prima est ab ortibus et occasibus stellarum fixarum desumpta. Nam si Terra annuo motu movetur, oportet latitudines ortivas et occiduas stellarum fixarum singulis 8 aut 10 diebus sensibiliter variari; sed consequens est falsum; ergo, et antecedens. Falsitas consequentis est nota: quia latitudines praedictae non variantur notabiliter nisi in 50 aut 60 annis. Consequentia vero probatur: quia, cum Terra simul cum horizonte moveatur sub zodiaco, et sic ab austro ad septentrionem et e contra, in 8 aut 10 diebus sensibiliter, fixae vero insensibiliter propter earum tardissimum motum sub zodiaco, imo secundum Copernicum sint immobiles, necesse est ut fixae ipsae in spatio 8 aut 10 dierum notabiliter suas latitudines ortivas et occiduas varient.*

*Altera est ab altitudinibus polaribus locorum. Nam si Terra movetur motu annuo, oportet mutari altitudines polares locorum; sed consequens est falsum; ergo, et antecedens. Falsitas consequentis est nota. Consequentia probatur: quia, cum Terra per annuum motum feratur a septentrione in austrum et e contra, simul etiam loca in ipsa existentia sic feruntur; ista autem latio mutat omnino altitudines polares: sicut enim homini qui a meridie ad boream vel e contra iter agit, contigit altitudinem poli mutari, ita loco continget mutari altitudinem poli, si vice hominis ipse moveatur.*

*Tertia est ab inaequalitate dierum artificialium. Nam, etiamsi videantur omnia observationibus consentire, dato motu Terrae annuo et Solis quiete, quia horizontis rectitudo seu obliquitas eadem semper existit, cum praesupponantur horizontes simul cum Terra moveri, tamen subtiliter intuenti non ita videbitur:*

*quia, cum per motum annuum transferatur Terra a borea in meridiem et e contra, necesse est ut zenith capitis nostri similiter transferatur, et ex consequenti ut aliquando accedat ad aequatorem et aliquando recedat; a mutatione autem zenith constat mutari rectitudinem et obliquitatem horizontis, quae inaequalitatem dierum potissimum efficit: ex quo consequitur, ut pro inaequalitate dierum signanda, non solum notandae essent differentiae motus annui, parallelos dierum artificialium efficientis, pro ut fit posito Solis motu et Terrae quiete, sed etiam illae differentiae quas efficerent mutationes obliquitatis horizontis dato motu Terrae et Solis quiete, et praecipue circa borealissimas habitationes, ubi variationes dierum sunt sensibilissimae: quod tamen non fit, cum solae primae differentiae animadvertantur, et illae observationibus satisfaciant.*

*Non obstat, quod horizontes simul cum Terra transferantur sine sui mutatione; quia verum esset hoc quoad motum Terrae diurnum, sed non quoad annuum: nam quoad hunc etiam quod transferantur cum Terra, tamen mutantur quoad obliquitatem et rectitudinem propter necessariam zenith mutationem, ut dictum est.*

*Quarta est ex Tychone in libro Epistolarum Astronomicarum, pag. 149, ubi asserit, cometas caelitus conspectos, et in Solis opposito versantes, motui Terrae annuo minime obnoxios esse, cum esse deberent, quia respectu ipsorum evanescere motum huiusmodi non est necesse, sicut in fixis syderibus, cum cometae praedicti illam maximam fixarum a Terra distantiam non habeant.*

*Contra denique tertium motum, Tycho in allegatis Epistolis obiicit tria: primo, quod sublato motu annuo, tertius necessario auferatur; 2°, quod fieri non potest ut axis Terrae in contrarium motui centri annuatim adeo correspondenter gyretur, ut quiescere tamen videatur; 3°, quod non potest dari in corpore unico et simplici axim et centrum duplici diversoque motu moveri; quibus si addatur etiam diurnus motus, maior efficitur difficultas.*

*ARGUMENTA PHYSICA CONTRA MOTUM TERRAE.*

*CAP. 6M.*

*Plurima possent adduci argumenta physica contra Terrae motionem, quae a philosophis et a mathematicis pro Terrae quiete afferuntur, et praecipue a Tychone in allegatis Epistolis: sed ego tria tantum in medium proponam. Quorum alterum est a natura corporum gravium et levium. Nam in universum videmus, corpora gravia minus apta esse ad motum quam levia aut non gravia: quod quidem statim innotescere potest consideranti non solum simplicia corpora naturalia, sed etiam mixta, et haec non solum in ordine ad motum qui a principio intrinseco causatur, sed etiam in ordine ad motum qui fit a principio extrinseco. Rursus videmus, naturam materias ita formis accommodare, ut pro efficientia ipsarum formarum miram animadvertamus ipsarum materierum aptitudinem: et id accidit tum quia, ut dicit Philosophus, 2° Physicorum, natura agit propter finem, tum quia materiae sunt velut instrumenta formarum ad agendum. Cum itaque Terra omnium corporum nostrae cognitioni subiectorum gravissima sit, oportet dicere naturam ei tot motus nequaquam tribuisse, et praecipue diurnum, adeo velocem ut in uno minuto temporis Terra conficere debeat fere 19 milliaria, ut dicit Tycho in Epistolis Astronomicis, pag. 190.*

*Alterum est quod desumitur ab illa physica propositione, unicuique corpori naturali unum esse tantummodo motum naturalem; quod verum esse, inductione probari facile posset, nisi ageretur cum philosopho praestantissimo. Cum itaque Terrae motus naturalis sit ad medium, non poterit ei esse naturalis motus circa medium, et multo minus poterunt ei esse naturales tot motus, et omnes non ad medium: si igitur motus illi Copernicei non sunt Terrae naturales, quomodo fieri potest ut Terra, corpus naturale, tamdiu illis moveatur? nam naturae non est praeter naturam agere.*

*Tertium est ab incongruentia quadam: quia scilicet omnibus caeli partibus lucidis, videlicet planetis, motum tribuit Copernicus; Soli autem, omnium coeli partium praestantissimo et lucidissimo, motum negat, ut Terrae, opaco et crasso*

*corpori, illum tribuat. Id enim facere non debuit, discretissima in omnibus suis operibus, natura.*

*ARGUMENTA THEOLOGICA CONTRA MOTUM TERRAE.*

*CAP. 7M.*

*Argumenta theologica ex Sacris Scripturis et authoritatibus Patrum et theologorum Scholasticorum infinita possent contra Terrae motionem proponi: sed duo tantum adducam, quae firmiora mihi esse videntur. Alterum est ex Iosue, cap. X, ubi ad preces Iosue dicit Scriptura: Stetit itaque Sol in medio coeli, et non festinavit occumbere spatio unius diei; non fuit antea et postea tam longa dies, obediente Domino voci hominis. Nec responsiones, quae afferuntur, quod Scriptura loquatur secundum modum nostrum intelligendi, satisfacit: tum quia in Sacris Litteris exponendis regula est ut semper litteralis sensus salvetur, cum fieri potest, ut in nostro casu; tum quia Patres omnes unanimiter exponunt locum hunc, quod Sol, qui movebatur, re vera stetit ad preces Iosue; ab ea vero interpretatione, quae est contra unanimem Patrum consensum, abhorret Tridentina Synodus, sess. 4th, in decreto de editione et usu Sacrorum Librorum, §* Praeterea. *Et licet Sancta Synodus loquatur in materia morum et Fidei, tamen negari non potest, quin Sanctis illis Patribus Sacrae Scripturae interpretatio contra consensum Patrum displiceat.*

*Alterum est ab authoritate Ecclesiae: nam in hymno ad vesperas feriae tertiae ita canit:*

> *Telluris ingens Conditor*
> *Mundi solum qui eruens,*
> *Pulsis aquae molestiis,*
> *Terram dedisti immobilem.*

*Nec leve est huiusmodi argumenti genus: nam, ut videre est apud Cardinalem Bellarminum, in plerisque locis confutat multos errores hymnis, canticis et precibus Ecclesiae, quae in breviariis habentur.*

*Et ex his absoluta sit haec disputatio. Cui respondere aut omnino aut ex parte, videlicet saltem mathematicis argumentis et physicis, et his non omnibus sed gravioribus, tuum arbitrium esto; nam hanc scripsi non ad tentandam eruditionem et doctrinam tuam, mihi omnibusque tum in Romana Curia tum extra notissimam, sed pro investigatione veritatis, quam te semper quaerere totis viribus profiteris: et re vera sic decet mathematicum ingenium.*

*FINIS*

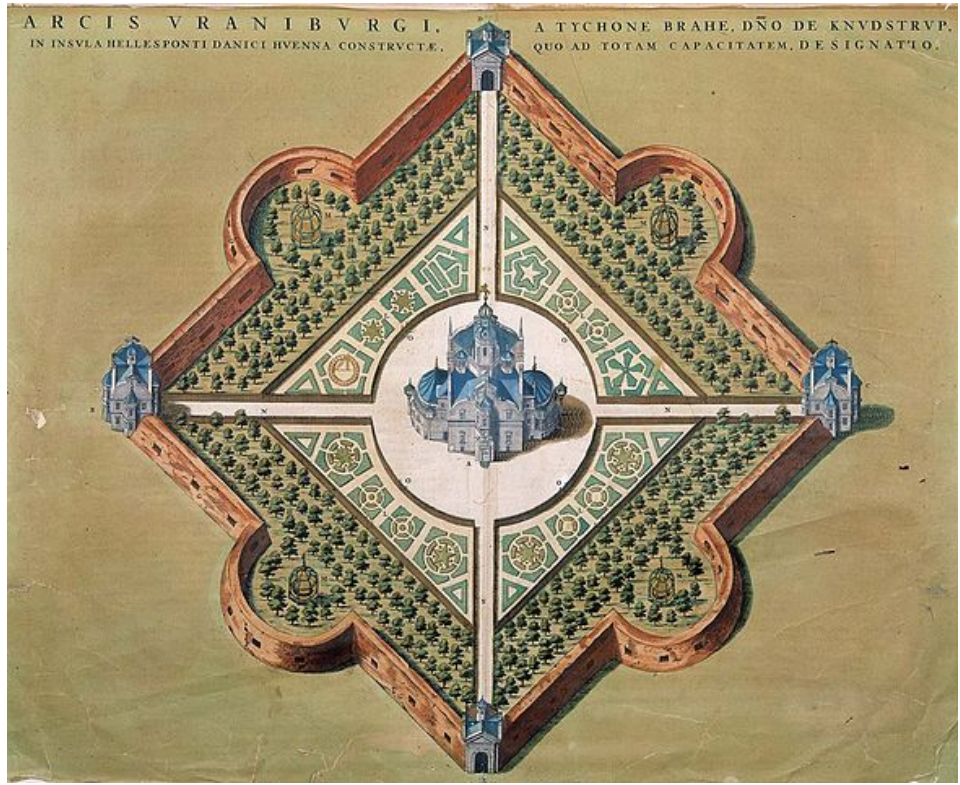
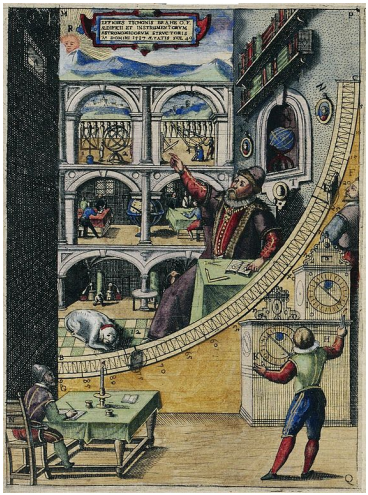
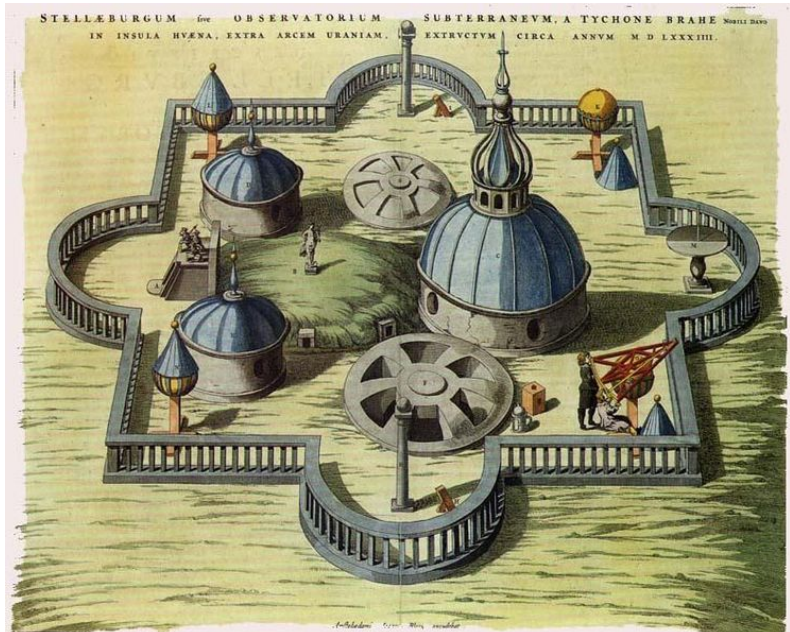

**FIGURE 1**

The facilities of Tycho Brahe's observatory.

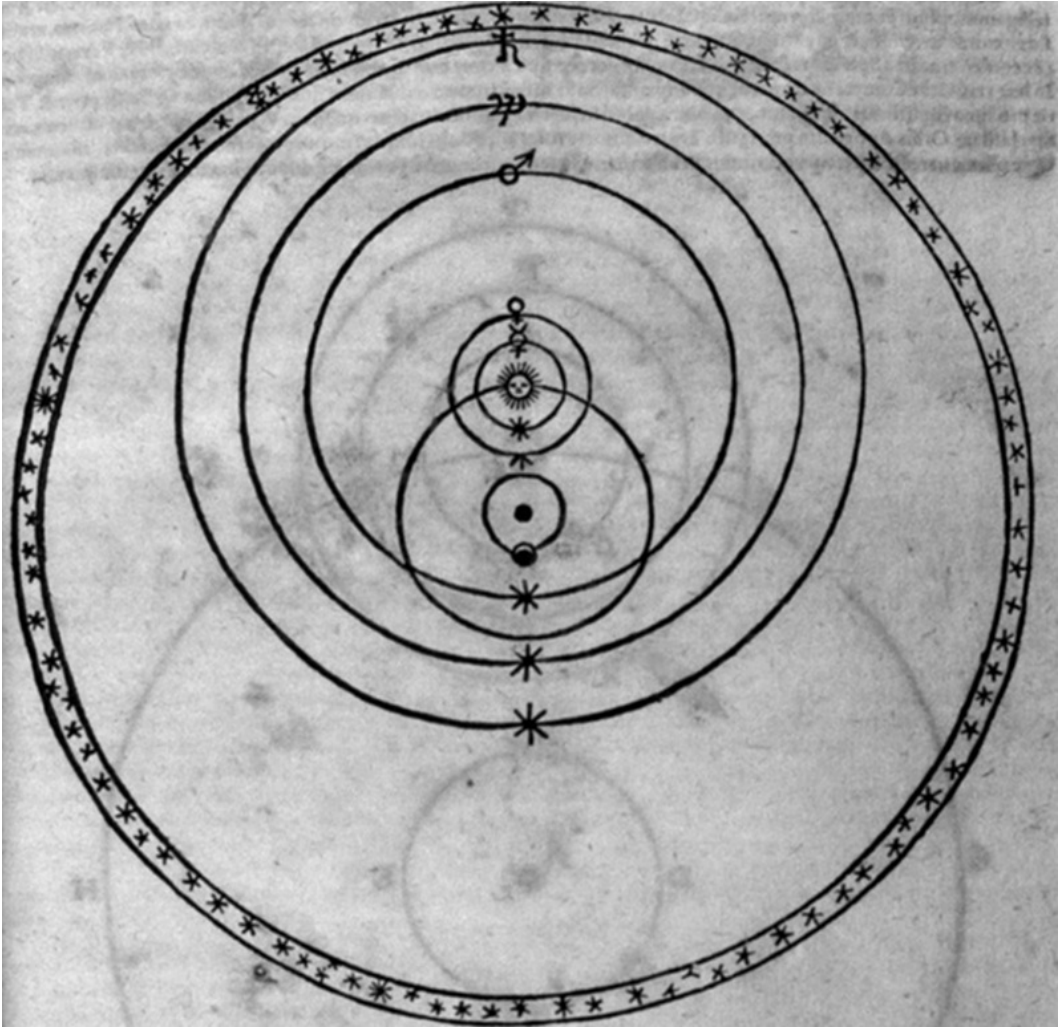

**FIGURE 2**
The Tychonic theory.  Note that the stars lie just beyond Saturn in this theory, in contrast to the Copernican theory in which the stars lie at vast distances from the Sun.

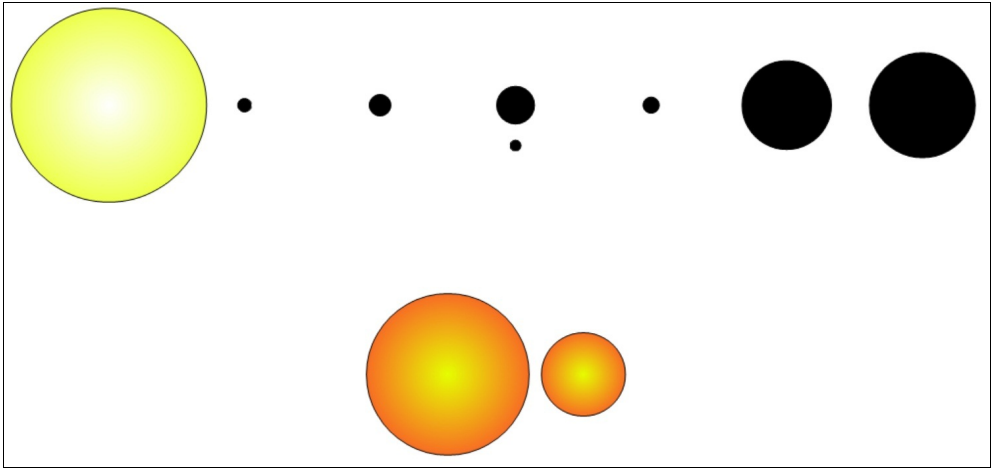

FIGURE 3

Above: Relative sizes of celestial bodies in a geocentric universe (in which the stars lie just beyond Saturn, as in Figure 2), calculated by Tycho Brahe, based on his observations and measurements.  From left to right (1$^{st}$ row) are the Sun, Mercury, Venus, Earth and Moon, Mars, Jupiter, Saturn, as well as (2$^{nd}$ row) a large star and a mid-sized star.

Below:  The figure from above (arrowed) compared to Brahe's calculated relative size for a mid-size star in the Copernican universe (in which the stars lie at vast distances, and thus must be enormous to explain their apparent sizes as seen from Earth).

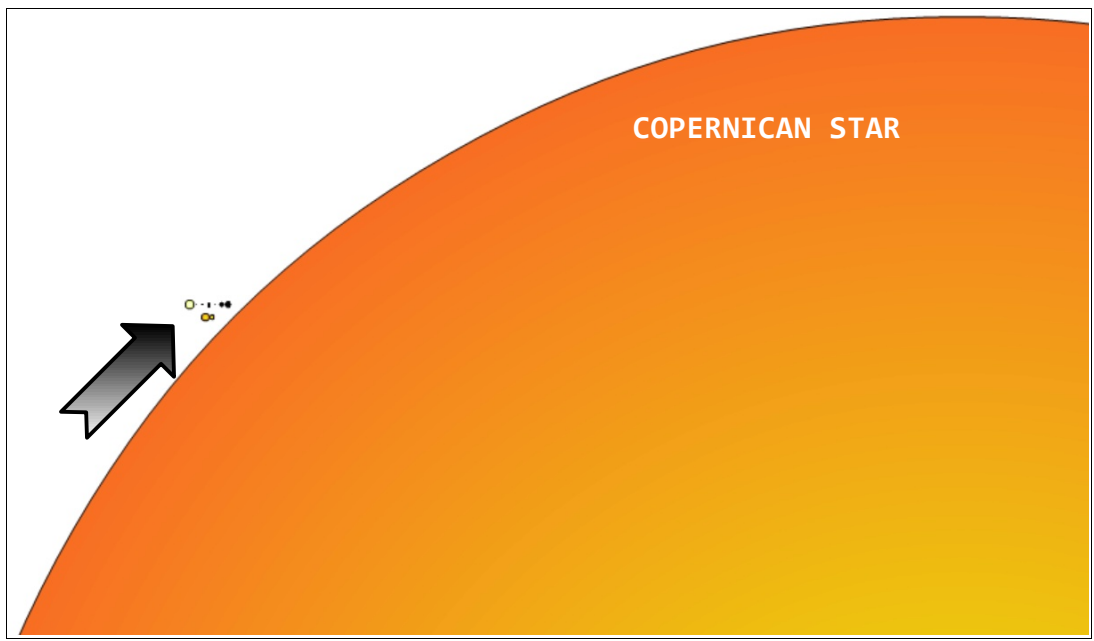

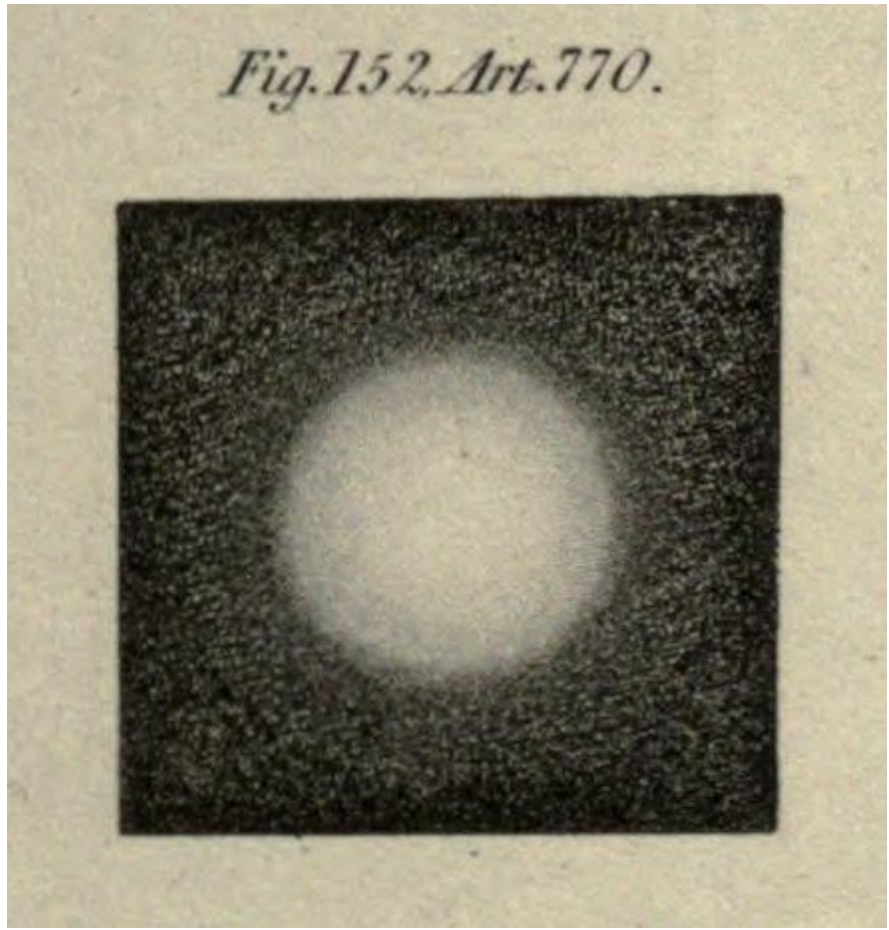

**FIGURE 4**

**Illustration of the globe-like appearance of a star seen through a small-aperture telescope, such as was used in the early seventeenth century. Early telescopic astronomers, including Galileo, Marius, and Riccioli, reported seeing the globes of stars in their telescopes. Since in the Copernican theory stars are an immense distance from Earth, the globe-like appearance translated into an immense physical size, far exceeding even that of the Sun; this apparently confirmed one of Tycho Brahe's key objections to the Copernican theory. The appearance is now known to be entirely illusory, caused by the diffraction of light passing through the telescope's aperture. The illustration is from the treatise on light written by the English astronomer John Herschel for the nineteenth-century *Encyclopædia Metropolitana*.**

Finocchiaro, Maurice A. 2005, *Retrying Galileo, 1633-1992* (University of California Press, Berkeley).

Finocchiaro, Maurice A. 2010, *Defending Copernicus and Galileo: Critical Reasoning in the Two Affairs* (Springer, Dordrecht).

Galilei, Galileo 1624, "Reply to Ingoli" in Finocchiaro, Maurice A. 1989, *The Galileo Affair: A Documentary History* (University of California Press: Berkeley).

Galilei, Galileo 2001 [1632], *Dialogue Concerning the Two Chief World Systems: Ptolemaic and Copernican, translated and with revised notes by Stillman Drake, foreword by Albert Einstein* (New York: The Modern Library).

Gingerich, Owen 1973, "Copernicus and Tycho", *Scientific American*, Vol. 229: 86-101.

Gingerich, Owen 1993, *The Eye of Heaven: Ptolemy, Copernicus, Kepler* (New York).

Gingerich, Owen 2009, "Galileo Opens the Door", Harvard-Smithsonian Center for Astrophysics 2009 Observatory Night Video Archive, February 19, 2009 <http://www.cfa.harvard.edu/events/mon_video_archive09.html>.

Gingerich, Owen & Voelkel, James R. 1998, "Tycho Brahe's Copernican Campaign", *Journal for the History of Astronomy*, Vol. 29: 1-34.